\documentclass{icrc2009}

\usepackage{graphicx}   
\usepackage[caption=false]{caption}    
\usepackage[font=footnotesize]{subfig} 
\usepackage{fixltx2e}
\usepackage{url}

\newcommand{\shorttitle}[1]%
{\markboth{Proceedings of the 31\MakeLowercase{$^{st}$} ICRC, {\L}\'{o}d\'{z} 2009}{#1} }
\newcommand{\etal}{\MakeLowercase{\textit{et al. }}} 

\def\Journal#1#2#3#4{{#4}, {#1}, {#2}, #3}

\def\ApJ{ApJ}

\begin{document}
\title{Discovery of Very High Energy $\gamma$-rays from the famous blazar S5~0716+714}

\author{\IEEEauthorblockN{Daniel Mazin\IEEEauthorrefmark{1},
			  Elina Lindfors\IEEEauthorrefmark{2},
                          Karsten Berger\IEEEauthorrefmark{3},
                          Nicola Galante\IEEEauthorrefmark{4},
                           Elisa Prandini\IEEEauthorrefmark{5} and \\
                           Takayuki Saito\IEEEauthorrefmark{4} 
                           for the MAGIC collaboration\IEEEauthorrefmark{6}}
                            \\
\IEEEauthorblockA{\IEEEauthorrefmark{1}IFAE, Edifici Cn., Campus UAB, E-08193 Bellaterra, Spain} 
\IEEEauthorblockA{\IEEEauthorrefmark{2}Tuorla Observatory, University of Turku, FI-21500 Piikki\"o, Finland}
\IEEEauthorblockA{\IEEEauthorrefmark{3}University of \L\'od\'z, PL-90236 Lodz, Poland}
\IEEEauthorblockA{\IEEEauthorrefmark{4}Max-Planck-Institut f\"ur Physik, D-80805 M\"unchen, Germany}
\IEEEauthorblockA{\IEEEauthorrefmark{5}Universit\`a di Padova and INFN, I-35131 Padova, Italy}
\IEEEauthorblockA{\IEEEauthorrefmark{6}\texttt{wwwmagic.mpp.mpg.de}}}

\shorttitle{Mazin \etal Discovery of VHE  $\gamma$-rays from S5~0716+714}
\maketitle

\begin{abstract}
The MAGIC collaboration reports the detection of the blazar S50716+714
(redshift uncertain) in very high energy gamma-rays. The observations were
performed in November 2007 and in April 2008, and were triggered by the KVA
telescope due to the high optical state of the object. The blazar S50716+714 is
the third low frequency BL Lac detected at energies above 100 GeV until today. 
Here, we present the results of the MAGIC observations.
\end{abstract}


\section{Introduction}

Blazars, a common term used for flat spectrum radio quasars (FSRQ) and
BL Lacertae objects, are the most extreme types of Active Galactic
Nuclei (AGN). In these objects the dominant radiation component
originates in a relativistic jet pointed nearly towards the
observer. 
The double-peaked spectral energy distribution (SED) of blazars is
attributed to a population of relativistic electrons spiraling in the
magnetic field of the jet. The low energy peak is due to synchrotron
emission and the second, high energy 
peak is attributed to inverse Compton scattering of
low energy photons in leptonic acceleration models. 
Models based on the acceleration of hadrons can also
sufficiently describe the observed SEDs and lightcurves \cite{Mannheim,
Mucke}. 
For most FSRQs and a 
large fraction of BL Lacertae objects (namely LBLs\footnote{LBL = low peak frequency
BL Lacertae}) the low energy peak is located in the energy range between
submillimeter to optical. On the other hand, for most of the sources detected
to emit VHE $\gamma$-rays (high frequency peaking BL Lacs, HBLs) the low energy peak is
located at UV to X-rays energies \cite{padovani07}. 

Blazars show variable flux in all wavebands from radio to
very high energy (VHE defined as $>$ 100 GeV) $\gamma$-rays. The
relationship between the variability at different wavebands appears
rather complicated. The MAGIC Collaboration is performing Target of
Opportunity observations of sources in a high flux state in the optical
and/or X-ray band. The optically triggered observations have resulted
in the discovery of VHE $\gamma$-rays from Mrk~180
\cite{mrk180} 
and 1ES~1011+496 \cite{1011}. 
In this paper we report the results of observations of the BL Lac object 
S5~0716+714 in November 2007 and April 2008, the latter resulting in the discovery
of VHE $\gamma$-rays from the source as announced in \cite{ATel}.

S5~0716+714 has been studied intensively at all
frequency bands. It is highly variable with rapid variations observed
from the radio to X-ray bands (Wagner et al. 1996). It has therefore
been target to several multiwavelength campaigns, the most recent one
organized by the GLAST-AGILE Support Program in July-November 2007
\cite{villata, giommi08a}.  
Due to the very bright nucleus, which outshines the host galaxy, the redshift 
of S5~0716+714 is still
uncertain. The recent detection of the host galaxy \cite{nilsson} 
suggests a redshift of $z = 0.31\pm0.08$ which is consistent
with the redshift $z = 0.26$ determined by spectroscopy for 3 galaxies
close to the location of S5~0716+714 \cite{stickel, bychova}. 
In the SED of S5~0716+714 the synchrotron peak in located 
in the optical band and is, therefore, classified either as
LBL \cite{nieppola}
or as IBL\footnote{IBL = intermidiate peak frequency BL Lacertae} \cite{padovani} 

S5~0716+714 was detected several times at different flux levels by
the EGRET detector on board the Compton Gamma-ray Observatory
\cite{3EG}. Also AGILE reported detection of a variable $\gamma$-ray
flux with peak flux density above the maximum reported from EGRET
\cite{AGILE}.
Observations at VHE $\gamma$-ray energies by HEGRA resulted in an upper limit of 
F($>1.6\,\mathrm{TeV})=3.13$ photons/cm$^2$/s \cite{Aharonian}. 
In this paper we present the first detection of VHE $\gamma$-rays from
S5~0716+714. It is the third optically triggered discovery of a new VHE
$\gamma$-ray emitting blazar by MAGIC. We also present MAGIC observations of the 
source in a low optical state, at which the observations resulted in a
flux upper limit that is clearly lower than the observed VHE $\gamma$-ray
flux during the optical high state. 


\section{Observations}
The MAGIC telescope is a stand alone imaging atmospheric Cherenkov telescope located on the Canary Island of La Palma. 
MAGIC has a standard trigger threshold of 60\,GeV, an angular
resolution of $ \sim 0.^\circ 1$ and an energy resolution
above 150\,GeV of $\sim 25\%$ (see \cite{crab} for details).

Tuorla blazar monitoring program \cite{takalo}  
(http://users.utu.fi/kani/1m) monitors S5~0716+714 on a nightly basis
using the KVA 35\,cm Telescope at La Palma and the Tuorla 1\,meter
telescopes. At the end of October 2007 (22th) the optical flux had
more than doubled (from 19\,mJy to 42\,mJy) in less than a month and MAGIC
was alerted. 
Due to moon and weather constraints, the
MAGIC observations started 11 days later, when the optical flux had
already decreased significantly (see Fig.~\ref{fig:lc}). MAGIC observed the
source for 14 nights for a total of 16.8 hours. During some nights the observing conditions were rather poor and the affected data was rejected from the
analysis. The observation time of the remaining good quality data amounts to 10.3 hours.
The zenith angle range of these observations was 42 to 46 degrees.

There was another strong optical flare from S5~0716+714 in April 2008. The
optical flux almost doubled within three nights (14th of April: 29 mJy, 17th April:
52 mJy), and at 17th of April reached its historical maximum value. MAGIC
started the observations 5 nights later, when the moon conditions allowed.
The source was observed during 9 nights with zenith angles from 47 to 55
degrees for a total of 7.1 hours. Unfortunately, during the last 6 nights of
the observations there was strong calima (sand from Sahara desert) in the
atmosphere and this data was, therefore, rejected from the analysis. 
The observation time of the remaining data of this period amounts to 2.8 hours
only.  
The total life time of S5~0716+714 MAGIC observations in 2007 and 
2008 after data quality cuts was 13.1 hours. 

In addition to R-band photometry we also made polarimetric observation
of S5~0716+714 using the KVA 60 cm telescope on 2008, April 29. 
The polarization was
$3.95\%\pm0.2\%$ and the polarization angle $102\pm2$. \cite{larinov} 
reported that between 2008, April 19 and April 25 the
positional angle of polarization rotated with an approximate rate 60
degrees per day and that after 2008, April 26 it returned to
the pre-outburst level. 
Since we measured the polarization only during one night,
we cannot confirm this behaviour of the positional angle of polarization.

\section{Data Analysis and Results}


\begin{figure}[t]
\centering
\includegraphics*[width=1.\columnwidth]{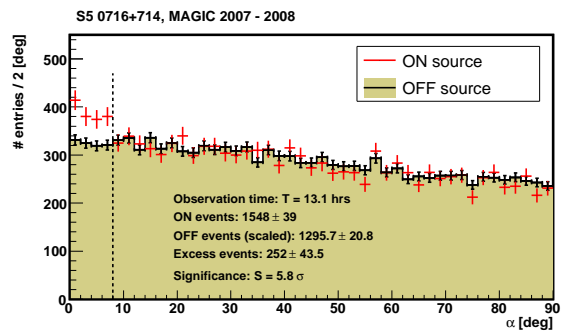}
\caption{\textsc{$\mid$Alpha$\mid$} distribution after all cuts for the total
MAGIC data sample in 2007 -- 2008. A $\gamma$-ray excess with a significance of
$5.8\,\sigma$ is found.} 
\label{fig:alpha} 
\end{figure}

\begin{figure}[t]
\centering
\includegraphics*[width=1.\columnwidth]{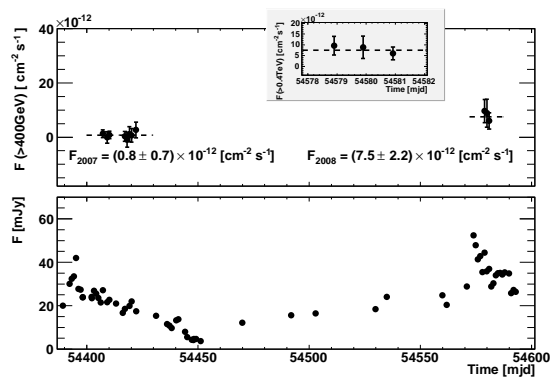}
\caption{The light curve of S5~0716+714 as measured from November 2007 until April 2008.
The day-by-day $\gamma$-ray light curve from MAGIC is shown in the upper panel whereas the optical 
KVA data is shown in the lower panel.
A zoom into the MAGIC 2008 data is shown in the inset of the upper panel to illustrate that there is no significant variability.}
\label{fig:lc} 
\end{figure}

The MAGIC data was analyzed using the standard analysis chain as described in 
\cite{NIMA,crab,timing}. 
Background rejection is achieved by a cut in \textsc{Hadronness}, which was optimized
using Crab Nebula data taken 
under comparable conditions (NSB noise, zenith distance, etc.).
The cut in \textsc{$\mid$Alpha$\mid$} that defines the signal region was also
optimized in the same way. An additional cut removed the events with a 
total charge of less than 200 photoelectrons (phe) in order to assure a better background rejection.
For the given cuts and zenith angle range of the observations the
analysis threshold corresponds to 400\,GeV.
The resulting \textsc{$\mid$Alpha$\mid$} distribution after all cuts 
for the overall S5~0716+714 data sample in 2007 -- 2008 is shown in 
Fig.~\ref{fig:alpha}. An overall excess of 252  $\gamma$-like
events over 1548 background events corresponding to a significance of $S =
5.8\,\sigma$ was found. Most of the signal comes from the 2008 data sample: the analysis of
the 2008 data only results in 176 excess events over 422 background events
corresponding to $S = 6.9\,\sigma$.  
From the 2007 data alone an excess corresponding to 
$S = 2.2\,\sigma$ was found.

\begin{figure}[t]
\centering
\includegraphics*[width=1.\columnwidth]{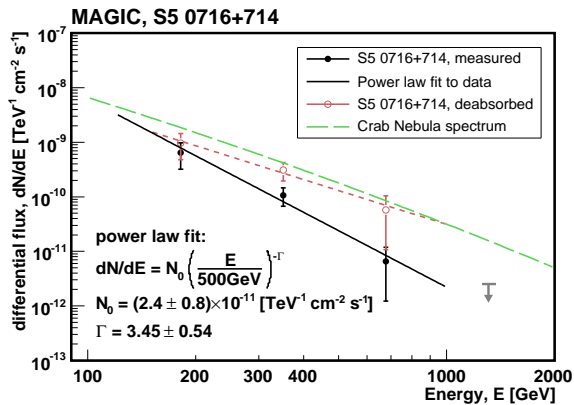}
\caption{The differential energy spectrum of S5~0716+714. Data from April 2008 are used only.
The measured (full black points) as well as deabsorbed (using the an EBL model
of \cite{primack05} and assuming z=0.31, hollow points)
spectra are shown. The highest energy point at 1.3\,TeV corresponds to a 95\% upper limit. 
The results from a power law fit to the measured spectrum are
shown in the plot.
The fitted power law to the deabsorbed spectrum has a photon index of $2.06 \pm 0.56$ (short--dashed line).
The Crab Nebula spectrum is shown for comparison (long--dashed line).} 
\label{fig:spec} 
\end{figure}


The day-by-day light curve as measured by MAGIC data is shown in Fig.~\ref{fig:lc} (upper
panel) together with the optical KVA light curve (lower panel).
In November 2007 the MAGIC flux above 400\,GeV is at 
$F_{\mathrm{2007}} (>0.4\mathrm{TeV}) = (0.8 \pm 0.7) [\mathrm{cm}^{-2} \mathrm{s}^{-1}]$,
whereas the flux is about 9 times higher in 2008:
$F_{\mathrm{2008}} (>0.4\mathrm{TeV}) = (7.5 \pm 2.2) [\mathrm{cm}^{-2} \mathrm{s}^{-1}]$.
No significant variability
is seen on time scales shorter than 6 months. However, we note that due to 
the observation windows and available data, we are not sensitive to week or months variability time scales.
No intranight variability was detected either but given the low significance of the individual MAGIC points, an intranight variability by a
factor of at least ten would have been required to detect it. In the optical band,
instead, a clear variability on time scales from days to months is visible with
two distinct flares: the first in October 2007, and the second in April 2008. 

The differential energy specrum is calculated only for the April 2008 data set.
The measured (and unfolded for detector effects) spectrum is shown in
Fig.~\ref{fig:spec}. The last energy point (at an energy of $E=1.3$\,TeV) has a
significance lower than 1\,$\sigma$ and was, therefore, converted into an
upper limit corresponding to a 95\% confidence level.  The measured spectrum
can be well fitted by a simple power law (with the differential flux
fiven in units of TeV$^{-1}$ cm$^{-2}$ s$^{-1}$):
\begin{equation}
\frac{\mathrm {d}N}{\mathrm{d}E\, \mathrm {d}A\, \mathrm {d}t} = (2.4\pm 0.8)\times10^{-11}(E/500\,\mathrm {GeV})^{-3.5\pm0.5} 
\end{equation}
Due to the energy-dependent attenuation of VHE $\gamma$-rays with low-energy
photons of the extragalactic background (EBL, \cite{gould}), the VHE
$\gamma$-ray flux of distant sources is significantly suppressed.  We
calculated the deabsorbed, i.e.\ intrinsic, spectrum of S5~0716+714 
using an EBL model of  \cite{primack05} and assuming
a redshift of $z = 0.31$. The resulting intrinsic spectrum (shown in
Fig.~\ref{fig:spec}, hollow points) has a fitted photon index of $\Gamma = 2.1
\pm 0.6$, which is well within the range of other extragalactic sources 
measured so far.

\section{Discussion}
MAGIC observed the blazar S5~0716+714 in November 2007 and April 2008, the
observations resulting in the discovery of a very high energy $\gamma$-ray
excess with a significance of $5.8\sigma$. During the November 2007
MAGIC observations the average optical flux was $\sim 20$mJy, while
during the optical flux was $\sim 45$mJy in April 2008. 
The same trend is also visible in the MAGIC data: the flux in
April 2008 is significantly higher than in November 2007. This seems
to support the indication seen in previous MAGIC observations for
other BL Lac objects \cite{mrk180,1011,bllac}, that there is
a connection between optical high states and VHE $\gamma$-ray high
states.  

In April 2008 S5~0716+714 was also in a historical high state in X-rays
\cite{giommi08b} and the optical polarization angle started to rotate
immediately after the optical maximum has been reached \cite{larinov}. 
However, the
radio flux at 37 GHz did remain at a quiescent level. This
multiwavelength behavior is very similar to the one in BL~Lac 2005
\cite{marscher}  indicating that the flaring events might have a 
similar origin. 

The SED modeling of the available data is not straight forward
as there are various simultaneous data suggesting a rather broad Inverse Compton
peak extending from high X-rays (measured by \textit{Swift}) to VHE $\gamma$-rays
(MAGIC). A detailed modeling of the source spectrum and 
an overall SED discussion will be published elsewhere.

As the source redshift is still uncertain, we used the MAGIC spectra
to calculate upper limits to the redshift. 
We assumed two different maximum possible 
hardness of the intrinsic spectrum: 
1.5, being a canonical value for a $\gamma$-ray spectrum emitted by electrons
with a spectral index of 2.0; and 0.666, being the limiting case
for  a $\gamma$-ray spectrum emitted by a monoenergetic electron distribution.
We get following upper limits
for the redshift: $z < 0.57$ (assuming the hardest intrinsic index of
1.5) and $z < 0.72$ (assuming the hardest intrinsic index of
2/3). Both limits agree with the redshift determined from the host
galaxy detection ($z = 0.31\pm0.08$) and 
from the spectroscopy of 3 nearby galaxies ($z = 0.26$).

%

\section*{Acknowledgment}

We would like to thank the Instituto de Astrofisica de 
Canarias for the excellent working conditions at the 
Observatorio del Roque de los Muchachos in La Palma. 
The support of the German BMBF and MPG, the Italian INFN 
and Spanish MICINN is gratefully acknowledged. 
This work was also supported by ETH Research Grant 
TH 34/043, by the Polish MNiSzW Grant N N203 390834, 
and by the YIP of the Helmholtz Gemeinschaft.
E.L.\ wishes to acknowledge the support by the Academy of Finland grant 127740.
D.M.'s research is supported by a Marie Curie Intra European Fellowship within 
the 7th European Community Framework Programme.

\end{document}